\documentclass[prc,twocolumn,superscriptaddress,showpacs,amssymb,amsmath,amsfonts,aps]{revtex4}
\usepackage{graphicx}
\usepackage{dcolumn}
\usepackage{epsfig}
\begin{document}

\title{
Extracting meson-baryon contributions to the electroexcitation 
of the $N(1675){\frac{5}{2}}^-$ nucleon resonance \\}

\newcommand*{\JLAB }{ Thomas Jefferson National Accelerator Facility,
Newport News, Virginia 23606, USA}
\affiliation{\JLAB }
\newcommand*{\YEREVAN }{ A.I.Alikhanian National Science Laboratory (Yerevan Physics Institute),
0036 Yerevan, Armenia}
\affiliation{\YEREVAN }
\author{I.G.~Aznauryan}
     \affiliation{\JLAB}
     \affiliation{\YEREVAN}
\author{V.D.~Burkert}
     \affiliation{\JLAB}
\date{\today}
\begin{abstract}
{We report on the determination of the electrocouplings for the transition from the 
proton to the $N(1675){\frac{5}{2}}^-$ resonance state using recent 
differential cross section data on $e p \rightarrow e\pi^+ n$ by the CLAS collaboration
at $1.8 \le Q^2 < 4.5$GeV$^2$. 
The data have been analyzed using two different 
approaches, the unitary isobar model and fixed-t dispersion relations.  
The extracted $\gamma^* p\rightarrow N(1675){\frac{5}{2}}^-$ helicity 
amplitudes show considerable 
coupling through the $A^p_{1/2}$ amplitude, that is significantly
larger than predicted three-quark contribution to this amplitude.
The amplitude $A^p_{3/2}$  
is much smaller. Both results are consistent with the predicted
sizes of the meson-baryon contributions at $Q^2 \geq 1.8~$GeV$^2$
from the dynamical coupled-channel model.} 
\end{abstract}
\pacs{ 13.40.Gp, 13.60.Le, 14.20.Gk, 25.30.Rw}
\maketitle

\section{Introduction}

The excitation of nucleon resonances has been the subject of great interest 
since the development of the quark model in 1964 \cite{GellMann1964,Zweig1964}.
The proposed 3-quark structure of the baryons  when realized in the dynamical 
quark models
resulted in prediction of a wealth of excited states  
with underlying spin-flavor and orbital symmetry of $SU(6) \otimes O(3)$. Most of 
the observed states have been found with hadronic probes, but they can also be 
investigated with electromagnetic probes \cite{Copley1969}. From the 
excited states predicted by the quark model, only a fraction has been observed to date.  
The search for the "missing" states  and more detailed studies of the resonance structure  
is now mostly carried out with electromagnetic probes and has been a major focus of  
hadron physics for the past decade \cite{Burkert2004}. This has led to a broad 
experimental effort in the development of large acceptance detectors and the measurement
of  exclusive meson photoproduction 
and electroproduction reactions, including many polarization observables. 
As a result, several new excited states 
of the nucleon have been discovered and entered into the 
Review of Particle Physics (RPP)~\cite{Agashe:2014kda}. Meson electroproduction  
revealed intriguing new information regarding the structure underlying the excited 
nucleon states \cite{Aznauryan2012}.  

One of the important insights is strong evidence that resonances are not excited 
from quark transitions alone, but there can be significant contributions from 
meson-baryon interactions 
as well, and that these two processes contribute to the excitation of the same state. 
This information has been obtained initially through the observation that the 
quark transition processes  often do not have sufficient strength 
to explain fully the measured transition amplitudes and form factors. One of the best known 
examples  is the photoexcitation of the $\Delta(1232){\frac{3}{2}}^+$ on the proton.
This  process is mostly 
due to a magnetic dipole transition from the nucleon, 
but only about 70\% of the magnetic dipole transition form factor at the real photon point
is explained by the quark content of the states.
A more satisfactory description of the 
$\gamma^* p\rightarrow \Delta(1232){\frac{3}{2}}^+$ 
transition was achieved in
models that incorporate pion-cloud contributions
\cite{Thomas,Faessler} and also in the
dynamical reaction models, where
the missing strength has been attributed  to 
dynamical meson-baryon interaction in the final state 
\cite{Kamalov1999,Kamalov2001,Sato2001,Sato2007,Sato2008}.

\begin{figure}
\begin{center}
\epsfig{file=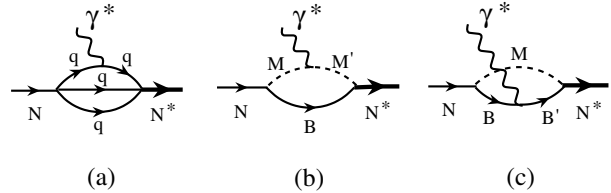,scale=0.54}
\caption{\small
Symbolic representation of the main contributions
to the  $\gamma^* N \rightarrow N^*$
transition: (a) through quark transition; (b,c) through meson-baryon
pairs.
\label{meson}}
\end{center}
\end{figure}

The two 
main processes that contribute to the  $\gamma^* N \rightarrow N^*$
transition
are illustrated in Fig. \ref{meson}
by the diagrams (a) and (b,c). 
The relative strength of these processes is determined
by the dynamics of the quark interaction and $SU(6)\otimes O(3)$
structure of the $N$ and $N^*$ in the diagram of Fig. \ref{meson}(a),
as well by the meson-baryon coupling constants and dynamics
of meson-baryon interaction in the diagrams of Figs. \ref{meson}(b,c).
The common feature of all approaches that account for meson-baryon
contributions is the fact that
these contributions are rapidly losing their strength when
$Q^2$ increases.
With electron scattering experiments, we have the tool to measure how the two
contributions change their relative strength as a function of the distance
scale, i.e. as we change $Q^2$.
Furthermore, at relatively small $Q^2$,
we have a unique handle to determine experimentally the non-quark
contribution to $\gamma^* N \rightarrow N^*$,
if excitation of the state through quark transition
is suppressed.
Most suitable for this purpose 
are measurements of the transition to resonant states with the 
total spin of the quarks $S_{3q} = {\frac{3}{2}}$ that belong to the 
$SU(6)\otimes O(3)$ multiplet $[70, 1^-]$.  
In the approximation of the single quark transition model (SQTM)
\cite{Hey_Weyers,Babcock_Rosner,Cottingham,SQTM}, 
for the excitation of these states on proton, both transverse helicity amplitudes 
are suppressed, i.e. $A^p_{1/2}=A^p_{3/2}=0$.
This suppression is known as the "Moorhouse" selection 
rule \cite{Moorhouse}, and is independent of $Q^2$.
The states with suppressed transverse amplitudes are 
$N(1650){\frac{1}{2}}^-$,  $N(1675){\frac{5}{2}}^-$,  
and $N(1700){\frac{3}{2}}^-$. 
Two of them, $N(1650){\frac{1}{2}}^-$ and $N(1700){\frac{3}{2}}^-$, have partners in the same 
multiplet with the same quantum numbers, $N(1535){\frac{1}{2}}^-$ and $N(1520){\frac{3}{2}}^-$, 
for which the quark contributions are not suppressed.
These states can mix.
The mixing angle for the $N(1650){\frac{1}{2}}^-$ is large, approximately $-31^\circ$ 
\cite{Isgur_Karl,Hey1975}, making it unsuitable as a candidate
for a measurement of the non-quark components. The second state, $N(1700){\frac{3}{2}}^-$,
 has a much smaller mixing angle of about $ +10^\circ$ and is a good candidate for such a measurement,
if it can be separated experimentally from the $\Delta(1700){\frac{3}{2}}^-$. 
While this separation is possible with the
data in at least two isospin configurations, e.g.  $ep\to e\pi^+ n$ and $ep\to e\pi^0 p$, 
the $\pi^0$ data do not exist now in the energy range needed 
for investigation of the $N(1700){\frac{3}{2}}^-$ and $\Delta(1700){\frac{3}{2}}^-$ contributions. 
This leaves the 
$N(1675){\frac{5}{2}}^-$ as the sole suitable candidate  
for a measurement of the non-quark
contributions to the transition amplitudes.  The suppression of the transverse amplitudes 
for this state is also obtained in dynamical quark
models that do not rely on the single quark approximation~\cite{Aznauryan_quark,Merten,Giannini}. 

\section{Data analysis and results}

The results on the 
$\gamma^* p \rightarrow~N(1675){\frac{5}{2}}^-$ transition reported 
in this paper have been extracted
as part of global fits to over 37,000 cross section 
data points collected recently with CLAS on the
$e p\rightarrow e n\pi^+$ 
at $1.8 \le   Q^2 < 4.5~$GeV$^2$ and $1.60 < W  < 2.0~$GeV \cite{Park}.
To further constrain the analysis, these data were combined
with the earlier CLAS data in the range $1.15 < W < 1.69~$GeV
at close values of $Q^2$ \cite{Park1}.
Therefore, the data sets at each $Q^2$ covered four resonance
regions from threshold to $2.0~$GeV.
Two conceptually different approaches, 
unitary isobar model (UIM)
and dispersion relations (DR), were utilized in the analysis
to model the non-resonant contributions
which must be separated from the direct s-channel resonance contributions. 
These approaches were described in detail in Refs.
\cite{Azn2003,Azn2009} and have been
successfully employed in Refs.
\cite{Azn2009,Azn065,Azn2005} for analyses of pion electroproduction data
in a wide range of $Q^2$ values from $0.16$ to $6~$GeV$^2$.

The UIM  \cite{Azn2003,Azn2009} was developed on the basis of MAID \cite{MAID}. 
At the values of $Q^2$ under investigation, the background
of the UIM \cite{Azn2003,Azn2009} is built from
the nucleon exchanges in the $s$-
and $u$-channels and $t$-channel
$\pi$, $\rho$ and $\omega$ exchanges.
This background is unitarized via unitarization of the multipole amplitudes
in the $K$-matrix approximation.
Resonance contributions are parametrized 
in the unified Breit-Wigner 
form with energy-dependent widths.

The DR approach \cite{Azn2003,Azn2009} is based on fixed-$t$ dispersion 
relations for invariant
amplitudes. They relate real parts of the amplitudes to 
the Born term ($s$- and $u$-channel nucleon and $t$-channel
$\pi$ exchanges) and integral over imaginary parts of the amplitudes. 
Taking into account isotopic structure,
there are 18 invariant amplitudes which describe $\pi$ electroproduction
on nucleons. For all these amplitudes, except one ($B_3^{(-)}$ in the notations
of Refs. \cite{Azn2003,Azn2009}), unsubtracted DR can be written.
For $B_3^{(-)}$, the subtraction is necessary. At the values of $Q^2$ 
under investigation, the subtraction was found empirically
in Ref. \cite{Azn2009} from the description of the data \cite{Park1}.
This subtraction was successfully employed in the present analysis.
In Ref. \cite{Azn2003}, the arguments were presented and discussed in detail, which show
that in $\pi$ electroproduction on nucleons, DR can be reliably used at $W \leq 1.8~$GeV.
The same conclusion was made
in early applications  of DR (see, for example, \cite{Crawford}).
Therefore, in our DR analysis, the
energy region is restricted by the first, second, and third resonance regions. 

\begin{figure*}
\begin{center}
\includegraphics[width=15.8cm]{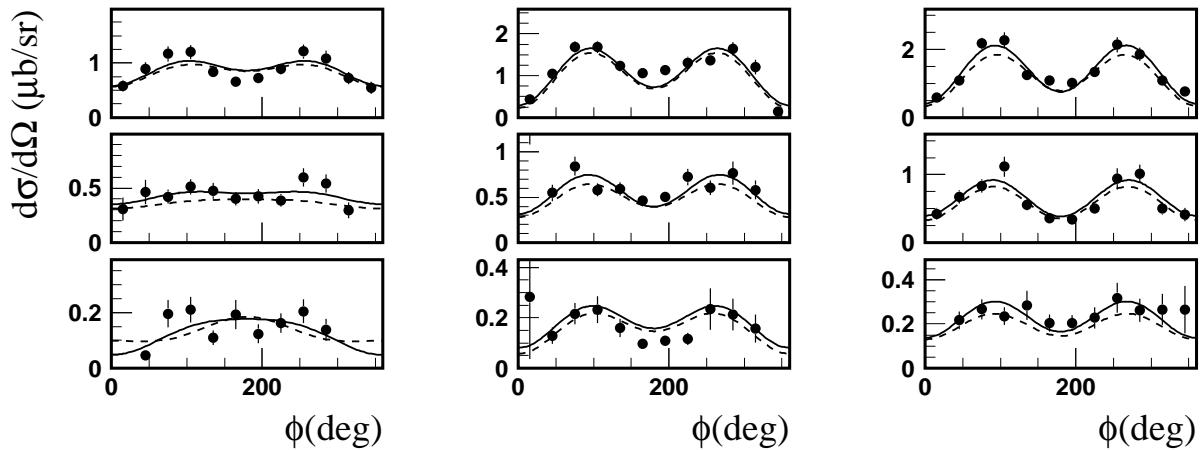}
\vspace{-0.1cm}
\caption{\small
Differential
cross sections
for the $\gamma^* p\rightarrow  n\pi^+$ reaction
at $W=1.675~$GeV
as a function of pion azimuthal angle $\phi$
at different values of pion polar angle $\theta$.
Data are from Ref. \cite{Park}.
The rows correspond to
$Q^2=1.8~$GeV$^2$, $\epsilon=0.866$;
$Q^2=2.6~$GeV$^2$, $\epsilon=0.792$;
$Q^2=4~$GeV$^2$, $\epsilon=0.626$.
The columns correspond to
$cos{\theta}=-0.5,~0.1,~0.5$.
The solid (dashed) curves
are the results obtained using UIM (DR) approach.
\label{str}}
\end{center}
\end{figure*}

Both approaches, UIM and DR, give comparable descriptions of the data
as is shown in Table~\ref{pip_data} and Fig.~\ref{str}.

\begin{table}
\begin{center}
\begin{tabular}{|cccccc|}
\hline
&&Number of&&$\chi^2/N$&\\
$Q^2$&$W$&data points&&&\\
(GeV$^2$)&(GeV)&($N$)&UIM&${}$&DR\\
\hline
1.72&1.15-1.69&3530&2.7&&2.9\\
1.8&1.6-2.01&8271&2.4&&\\
&1.6-1.8&5602&2.3&&2.4\\
2.05&1.15-1.69&5123&2.3&&2.5\\
2.2&1.6-2.01&8140&2.2&&\\
&1.6-1.8&5539&2.3&&2.3\\
2.44&1.15-1.69&5452&2.0&&2.3\\
2.6&1.6-2.01&7819&1.7&&\\
&1.6-1.8&5373&2.0&&2.2\\
2.91&1.15-1.69&5484&2.1&&2.3\\
3.15&1.6-2.01&7507&1.8&&\\
&1.6-1.8&5333&2.1&&2.0\\
4.16&1.15-1.69&5778&1.2&&1.3\\
4.0&1.6-2.01&5543&1.3&&\\
&1.6-1.8&4410&1.5&&1.6\\
\hline
\end{tabular}
\caption{\label{pip_data}
The values of $\chi^2$ for the 
$\gamma^* p \rightarrow \pi^+ n$ cross sections
obtained in the analyses within UIM and DR.
The data at $Q^2=1.8,2.2,2.6,3.15,4$GeV$^2$ and 
$Q^2=1.72,2.05,2.44,2.91,4.16$GeV$^2$ are, respectively, from 
Refs. \cite{Park,Park1}.}
\end{center}
\end{table}

In the global analysis, we have taken into account
all  4- and 3-star resonances
from the first, second, and third resonance
regions: 
$\Delta(1232){\frac{3}{2}}^+$, $N(1440){\frac{1}{2}}^+$,
$N(1520){\frac{3}{2}}^-$, $N(1535){\frac{1}{2}}^-$,
$\Delta(1600){\frac{3}{2}}^+$, $\Delta(1620){\frac{1}{2}}^-$,
$N(1650){\frac{1}{2}}^-$,
$N(1675){\frac{5}{2}}^-$,
$N(1680){\frac{5}{2}}^+$,
$N(1700){\frac{3}{2}}^-$, $\Delta(1700){\frac{3}{2}}^-$,
$N(1710){\frac{1}{2}}^+$, and
$N(1720){\frac{3}{2}}^+$.
From the resonances of fourth resonance region,
we have included the resonances
$\Delta(1905){\frac{5}{2}}^+$ and
$\Delta(1950){\frac{7}{2}}^+$ which have been clearly seen in
$\pi$ photoproduction.
For the masses, widths, and  $\pi N$ branching
ratios  of the resonances we used
the mean values of the data from the 
RPP~\cite{Agashe:2014kda}. 
The results on the resonances of
the first and second resonance regions including their model uncertainties
are based on the data \cite{Park1}.
They have been found and presented
in Ref. \cite{Azn2009}.  
The analysis of the combined sets of data
\cite{Park,Park1} allowed us to get reliable results
for the electroexcitation amplitudes of the states  
$N(1675){\frac{5}{2}}^-$,
$N(1680){\frac{5}{2}}^+$, and $N(1710){\frac{1}{2}}^+$.
The isotopic pairs of the resonances 
$\Delta(1600){\frac{3}{2}}^+$ and $N(1720){\frac{3}{2}}^+$.
$\Delta(1620){\frac{1}{2}}^-$ and $N(1650){\frac{1}{2}}^-$,
$\Delta(1700){\frac{3}{2}}^-$ and $N(1700){\frac{3}{2}}^-$
could not be separated from each other from the data on the $N\pi$ production
in one channel. For their investigation, data in at least two
channels, e.g. 
$\gamma^* p\rightarrow n \pi^+$ and $\gamma^* p\rightarrow p \pi^0$, are necessary.

Here we present and discuss the results on the
$N(1675){\frac{5}{2}}^-$, because of the unique role the state plays in the
study of the meson-baryon contributions to the $\gamma^* N \rightarrow N^*$
transition amplitudes.
Detailed results on the CLAS data and their global analysis 
have been presented in Ref.~\cite{Park}.

\begin{figure*}
\begin{center}
\vspace{-0.3cm}
\includegraphics[width=12.8cm]{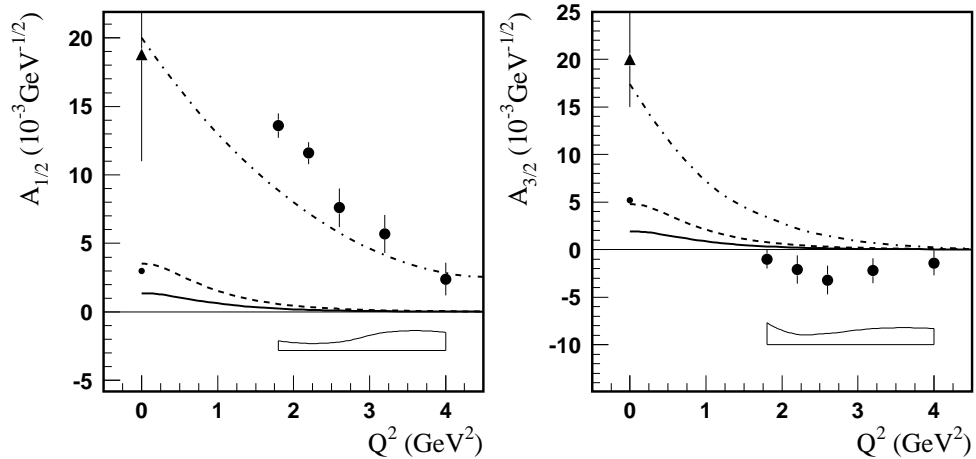}
\vspace{-0.3cm}
\caption{\small
Transverse helicity amplitudes
for the \protect $\gamma^* p \rightarrow~N(1675){\frac{5}{2}}^-$
transition. 
The full circles are 
the amplitudes extracted from the data \cite{Park,Park1}
using the following mass, width, and $\pi N$ branching
ratio of the resonance:
$M=1.675~$GeV, $\Gamma=0.15~$GeV, and $\beta_{\pi N}=0.4$.
The bands show the model uncertainties.
The full triangles at $Q^2=0$ are the RPP estimates \cite{Agashe:2014kda}.
The solid and dashed curves correspond
to quark model predictions of Refs. \cite{Giannini} and
\cite{Merten}, respectively; 
the dots at $Q^2=0$ are the predictions of the light-front
relativistic quark model from Ref. \cite{Aznauryan_quark}.
Dashed-dotted curves are absolute values of the
predicted meson-baryon contributions
from the dynamical coupled-channel model of Ref. \cite{Sato2008}.
\label{d15}}
\end{center}
\end{figure*}

The results 
for the  $\gamma^* p \rightarrow~N(1675){\frac{5}{2}}^-$
transverse helicity amplitudes extracted from the
experimental data 
are shown in Fig. \ref{d15}. 
The presented amplitudes 
are averaged values of the results obtained using UIM and DR.
The uncertainty that originates from the averaging is considered
as one of the model uncertainties. 
We consider also two other kinds
of model uncertainties.
The first one arises from
the uncertainties of the widths and masses of the resonances.
The second one 
is related to the uncertainties of the background of UIM
and the Born term in DR.
The pion and nucleon electromagnetic form factors
that enter these quantities are known quite well
from experimental data \cite{Melnitchouk,Lachniet,Horn,Tadevos,Riordan},
and the second uncertainty is caused mainly by the poor knowledge of
the $\rho \rightarrow \pi \gamma$ form factor.
According to the QCD sum rule \cite{Eletski}
and quark model \cite{AznOgan} predictions, the $Q^2$ dependence
of this form factor is
close to the dipole form
$G_D(Q^2)
=1/(1+\frac{Q^2}{0.71GeV^2})^2$.
We used this form in our analysis
and have introduced in our final results a systematic
uncertainty that accounts for a $20\%$ deviation
from $0.71~$GeV$^2$.
All these uncertainties added in quadrature are presented as model
uncertainties of the amplitudes. 

In Fig. \ref{d15}, we show also the predictions of different quark models
\cite{Aznauryan_quark,Merten,Giannini} that do not account for  
meson-baryon contributions.
They are consistent with each other and confirm
the strong suppression of the  
$\gamma^* p \rightarrow~N(1675){\frac{5}{2}}^-$ transverse helicity amplitudes that
follows from the SQTM \cite{SQTM}. 
The values of $A^p_{1/2}$ and $A^p_{3/2}$, predicted by quark models,
are smaller than statistical and model uncertainties of the
amplitudes extracted from the data \cite{Park,Park1}. 
Therefore, taking into account these uncertainties,  
the experimental $\gamma^* p \rightarrow~N(1675){\frac{5}{2}}^-$ amplitudes  
can be considered as determined predominantly by the contributions
caused by meson-baryon effects. 
The extracted amplitudes show significant
coupling through $A^p_{1/2}$,
while $A^p_{3/2}$
is consistent with 0 within statistical and model uncertainties.
Taking into account values of the amplitudes at $Q^2=0$ {\cite{Agashe:2014kda}},
we conclude that $A^p_{3/2}$ drops much faster than $A^p_{1/2}$.

In Fig. \ref{d15}, we show the results
from the dynamical coupled-channel approach by the
EBAC group \cite{Sato2008}. In this approach the meson-baryon
contributions to the 
$\gamma^* p \rightarrow~N(1675){\frac{5}{2}}^-$ amplitudes have been
found at the resonance pole position.
The amplitudes that are presented  
in Fig. \ref{d15} are absolute values of these contributions continued
to the real axis close to the mass of the 
state $N(1675){\frac{5}{2}}^-$. 
$Q^2$-behaviour of the amplitudes extracted from experiment and their sizes 
are qualitatively 
consistent  with these results of the dynamical coupled-channel approach \cite{Sato2008}. 

\section{Summary}

Based on new high precision data from CLAS in the 
$ep\to e\pi^+n$ channel \cite{Park}, combined with previously obtained 
data on the same channel at close values of $Q^2$ but at lower values of $W$~\cite{Park1}, 
we have extracted the electroexcitation helicity amplitudes for the 
resonance $N(1675){\frac{5}{2}}^-$. A special feature 
of this state is the strong suppression of the transverse helicity amplitudes for 
its excitation through quark transition from the proton. This feature 
allowed us to draw conclusions regarding the dominant strength 
of the meson-baryon contributions to the $\gamma^* p \rightarrow~N(1675){\frac{5}{2}}^-$ transition. 
The results are important as unambiguous experimental test  for  
models that account for the meson-baryon contributions to the
electroexcitation of nucleon resonances and will support theoretical developments
towards a more complete understanding of the dynamics
of nucleon resonance excitations. 

The data \cite{Park,Park1} cover the relatively high $Q^2$ range.
It would be desirable to study the state $N(1675){\frac{5}{2}}^-$
at lower $Q^2$ to map out the transition to the real photon point
where the amplitudes are not well known. Furthermore, measurements
on the neutron are very desirable, as significant strength
through quark transition is expected for both transverse amplitudes
in the excitation of the $N(1675){\frac{5}{2}}^-$ from the neutron \cite{SQTM}.

{\bf Acknowledgments}. This work was supported by the
State Committee of Science of Republic of Armenia, Grant 13-1C023,
and the US Department of Energy, Office of Science,
Office of Nuclear Science, under contract No. DE-AC05-06OR23177.


\begin{thebibliography}{999}


\bibitem{GellMann1964} 
  M. Gell-Mann,
  Phys. Lett.  {\bf 8}, 214 (1964).

\bibitem{Zweig1964} 
  G. Zweig, CERN Report Nos. TH 401 and 412 (1964).

\bibitem{Copley1969} 
  L. A. Copley, G. Karl and E. Obryk,
  Nucl. Phys. B {\bf 13}, 303 (1969).

\bibitem{Burkert2004}
  V. D. Burkert and T.-S. H. Lee,
  Int. J. Mod. Phys. E {\bf 13}, 1035 (2004).
  
\bibitem{Agashe:2014kda} 
  K.~A.~Olive et al.  (Particle Data Group),
  Chin. Phys. C {\bf 38}, 090001 (2014).  
  
\bibitem{Aznauryan2012}
  I. G. Aznauryan and V. D. Burkert,
  Prog. Part. Nucl. Phys. {\bf 67}, 1 (2012).
  
\bibitem{Thomas} D. H. Lu, A. W. Thomas, and A. G. Williams,
Phys. Rev. C {\bf 55}, 3108 (1997).

\bibitem{Faessler} A. Faessler, T. Gutsche, B.R. Holstein et al.,
Phys. Rev. D {\bf 74}, 074010 (2006).

\bibitem{Kamalov1999} S.S. Kamalov and S.N. Yang,
Phys. Rev. Lett. {\bf 83}, 4494 (1999).

\bibitem{Kamalov2001} S.S. Kamalov et al.,
Phys. Rev. C {\bf 64} (R), 032201 (2001).

\bibitem{Sato2001} T. Sato and T.-S. H. Lee,
Phys. Rev. C {\bf 63}, 055201 (2001).

\bibitem{Sato2007}  A. Matsuyama, T. Sato, and T.-S.H. Lee,
Phys. Rep. {\bf 439}, 193 (2007).

\bibitem{Sato2008} B. Juli\'a-D\'iaz, T.-S.H. Lee, A. Matsuyama, and T. Sato,
Phys. Rev. C {\bf 77}, 045205 (2008).

\bibitem{Hey_Weyers}
A.J.G. Hey and J. Weyers,
Phys. Lett. B {\bf 48}, 69 (1974).

\bibitem{Babcock_Rosner}
J. Babcock and J.L. Rosner,
Ann. Phys. (N.Y.)  {\bf 96}, 191  (1976) .

\bibitem{Cottingham}
W.N. Cottingham and I.H. Dunbar,
Z. Phys.  C {\bf 2}, 41 (1979). 

\bibitem{SQTM}
V.D. Burkert et al.,
Phys. Rev.  C {\bf 67}, 035204 (2003). 

\bibitem{Moorhouse}
R. G. Moorhouse,
Phys. Rev. Lett. {\bf 16}, 772 (1966).

\bibitem{Isgur_Karl}
N. Isgur and G. Karl,
Phys. Lett. B {\bf 72}, 109 (1977).

\bibitem{Hey1975}
A.J.G. Hey, P.J. Litchfield, and R.J. Cashmore,
Nucl. Phys. B {\bf 95}, 516 (1975).

\bibitem{Aznauryan_quark} I. G. Aznauryan and A. S. Bagdasaryan, Yad. Fiz. 
$\bf{41}$, 249 (1985); translation in Sov. J. Nucl. Phys.
$\bf{41}$, 158 (1985). 

\bibitem{Giannini} E. Santopinto and M. M. Giannini,
Phys. Rev. C {\bf 86}, 065202 (2012).

\bibitem{Merten} D. Merten, U. L\"oring, 
B. Metsch, and H. Petry,
Eur. Phys. J. A {\bf 18}, 193 (2003).

\bibitem{Park} K. Park et al., CLAS Collaboration, arXiv:1412.0274.

\bibitem{Park1} K. Park et al., CLAS Collaboration,
Phys. Rev. C {\bf 77}, 015208 (2008).

\bibitem{Azn2003} I. G. Aznauryan,
Phys. Rev. C {\bf 67}, 015209 (2003).

\bibitem{Azn2009} I. G. Aznauryan et al., CLAS Collaboration,
Phys. Rev. C {\bf 80}, 055203 (2009).

\bibitem{Azn2005} I. G. Aznauryan, V. D. Burkert, H. Egiyan,
et al., Phys. Rev. C {\bf 71}, 015201 (2005).

\bibitem{Azn065} I. G. Aznauryan, V. D. Burkert,
et al., Phys. Rev. C {\bf 72}, 045201 (2005).

\bibitem{MAID} D. Drechsel, O. Hanstein, S. Kamalov,  and L. Tiator,
Nucl. Phys. A $\bf{645}$, 145 (1999).

\bibitem{Crawford}  R. L. Crawford,
Nucl. Phys. B $\bf{97}$, 125 (1975).

\bibitem{Melnitchouk} J. Arrington, W. Melnitchouk,
J. A. Tjon,
Phys. Rev. C {\bf 76}, 035205 (2007).

\bibitem{Lachniet} J. Lachniet et al., CLAS
Collaboration, Phys. Rev. Lett. {\bf 102}, 192001 (2009).

\bibitem{Horn} T. Horn  et al., Phys. Rev. Lett. {\bf 97}, 192001
(2006).

\bibitem{Tadevos} V. Tadevosyan et al., Phys. Rev. C  {\bf 75},
055205 (2007).

\bibitem{Riordan} S.Riordan et al., Phys.Rev.Lett. 
{\bf 105}, 262302 (2010).

\bibitem{Eletski} V. Eletski and Ya. Kogan, Yad. Fiz. {\bf 39}, 138
(1984).

\bibitem{AznOgan} I. Aznauryan and K. Oganessyan, Phys. Lett. B
{\bf 249}, 309 (1990).

\end{thebibliography}
\end{document}